\documentclass[aps,prl,10pt,twocolumn,floatfix,superscriptaddress,showpacs,numerical,footinbib,longbibliography]{revtex4-1}
\usepackage[normalem]{ulem}
\usepackage{graphicx}
\usepackage{amsmath}
\usepackage{amssymb}
\usepackage{mathdots}
\usepackage{bbold}
\usepackage[usenames,dvipsnames,pdftex]{color}
\usepackage[%
  colorlinks=true,
  urlcolor=blue,
  linkcolor=blue,
  citecolor=blue
]{hyperref}

\begin{document}

\title{Spin-1 Kitaev-Heisenberg model on a two-dimensional honeycomb lattice}

\author{Xiao-Yu Dong }
\affiliation{Department of Physics and Astronomy, California State University, Northridge, CA 91330, USA}
\affiliation{Department of Physics and Astronomy, Ghent University, Krijgslaan 281, 9000 Gent, Belgium}
\author{D. N. Sheng}
\affiliation{Department of Physics and Astronomy, California State University, Northridge, CA 91330, USA}

\begin{abstract}
We study the Kitaev-Heisenberg model with spin-1 local degree of freedom on a two-dimensional honeycomb lattice numerically by density matrix renormalization group method. By tuning the relative value of the Kitaev and Heisenberg exchange couplings, we obtain the whole phase diagram with two spin liquid phases and four symmetry broken phases. We identify that the spin liquid phases are gapless by calculating the central charge at the pure Kitaev points without Heisenberg interaction. Comparing to its spin-1/2 counterpart, the position and number of gapless modes of the spin-1 case are quite different. Due to the approximate $Z_2$ local conservations, the expectation value of Wilson loop operator measuring the flux of each plaquette stays near to 1, and the static spin-spin correlations remain short-range in the entire spin liquid phases. 
\end{abstract}

\maketitle
\textbf{Introduction.}
Ever since it was first proposed by Alexei Kitaev in 2006~\cite{kitaev2006anyons}, the two-dimensional (2D) Kitaev model and its various extensions have drawn extensive attention from both theoretical and experimental side~\cite{hermanns2018physics,rau2016spin}. The Kitaev model with spin-1/2 local degree of freedom on a honeycomb lattice could be solved exactly by representing the spin operators with Majorana fermions in extended space, and then the system decouples into a free Majorana fermion system and a static $Z_2$ gauge field. Depending on the relative relations between the exchange couplings in different spatial directions, the system could be a gapless spin liquid or a gapped $Z_2$ topologically ordered spin liquid. Under a small external magnetic field, the gapless spin liquid will open a gap and change to a non-Abelian topological phase~\cite{zhu2018robust, zou2018field,jiang2011possible,janssen2016honeycomb,gohlke2018dynamical}, while under a strong magnetic field, the system will be in a fully polarized phase with chiral magnon edge states~\cite{mcclarty2018topological}. The existence of a U(1) gapless quantum spin liquid phase is also proposed in an intermediate magnetic field~\cite{jiang2018field,PhysRevB.100.165123,liang2018intermediate,hickey2019emergence}. The other extensions, for example, the addition of Heisenberg exchange terms, may lead to different magnetically ordered phases with spontaneous symmetry breaking~\cite{iregui2014probing, gohlke2017dynamics,chaloupka2010kitaev,sela2014order}. Some candidate materials with signatures of Kitaev physics have also been found~\cite{trebst2017kitaev,rau2016spin,winter2017models,jackeli2009mott}, such as, $\alpha$-RuCl$_3$~\cite{kasahara2018majorana,plumb2014alpha,banerjee2016proximate,winter2017breakdown,little2017antiferromagnetic,wang2017magnetic,ran2017spin}, H$_3$LiIr$_2$O$_6$~\cite{bette2017solution}, and A$_2$IrO$_3$(A = Na, Li)~\cite{chaloupka2013zigzag,chun2015direct,choi2012spin,singh2010antiferromagnetic,singh2012relevance,katukuri2014kitaev,chun2015direct,williams2016incommensurate,rau2014generic,chaloupka2010kitaev}, which motivate more studies on various extended Kitaev models with 
additional couplings as  the effective Hamiltonians for reaslistic materials.

Despite the numerous studies on spin-1/2 Kitaev model, the spin-1 case is still less explored and has not been fully understood. The examples of spin-1 spin liquid are quite rare in Heisenberg-like spin models, and the Hamiltonian terms giving rise to frustration need to be strong enough to overcome the tendencies of magnetic ordering or other spontaneous symmetry breaking. 
However, with increasing of local spin, quantum fluctuation is reduced while the Hilbert space is enlarged. This combined effect makes spin-1 models interesting and the underlying physics of spin-1 systems may be richer.
Most of the current understandings of spin-1 Kitaev model are based on the symmetry analysis, such as the existence of $Z_2$ Wilson loop operators and the vanishing of spin-spin correlations beyond nearest neighbors~\cite{baskaran2008spin, baskaran2007exact}. 
Small sizes numerical calculations~\cite{koga2018ground} using exact diagonalization suggested that the ground state of the spin-1 Kitaev model is possibly gapless and vortex-free.
Unlike the spin-1/2 Kitaev model,  there is no exact solution to such a model,
and   the nature of the quantum state is far from understood.
A recent theoretical paper~\cite{stavropoulos2019microscopic} proposed the 
relevance of  the spin-1 Kitaev spin model to 2D Mott insulators with strong
spin-orbit coupling on lattices with edge-shared octahedra, which may be realized in layered transition metal oxides. This work opens the possibility to find spin-1 spin liquid phase in real materials and makes the theoretical understanding of spin-1 Kitaev model more urgent and important.

In this work, we numerically study the spin-1 Kitaev-Heisenberg model on a 2D honeycomb lattice using density matrix renormalization group (DMRG) method. We obtain the whole phase diagram with different relative values of the Kitaev and Heisenberg exchange couplings. Two spin liquid phases exist around the anti-ferromagnetic and ferromagnetic pure Kitaev points (with no Heisenberg term), respectively. The other four phases are magnetically ordered with different patterns of spontaneous symmetry breaking. The entanglement spectrum of the spin liquid phases has a degeneracy with different momentum around a cylinder, while the spectrum of the magnetically ordered phases is dominated by the largest Schmidt value. 
Furthermore, we find that the ground states of the pure Kitaev points are gapless with a non-zero central charge. However, the position and number of the gapless modes are different from its spin-1/2 counterpart. The expectation value of Wilson loop operator measuring flux of each plaquette stays near to 1, and the static spin-spin correlations remain short-range in the entire spin liquid phases.

\begin{figure}[h]
 \includegraphics[width=\columnwidth]{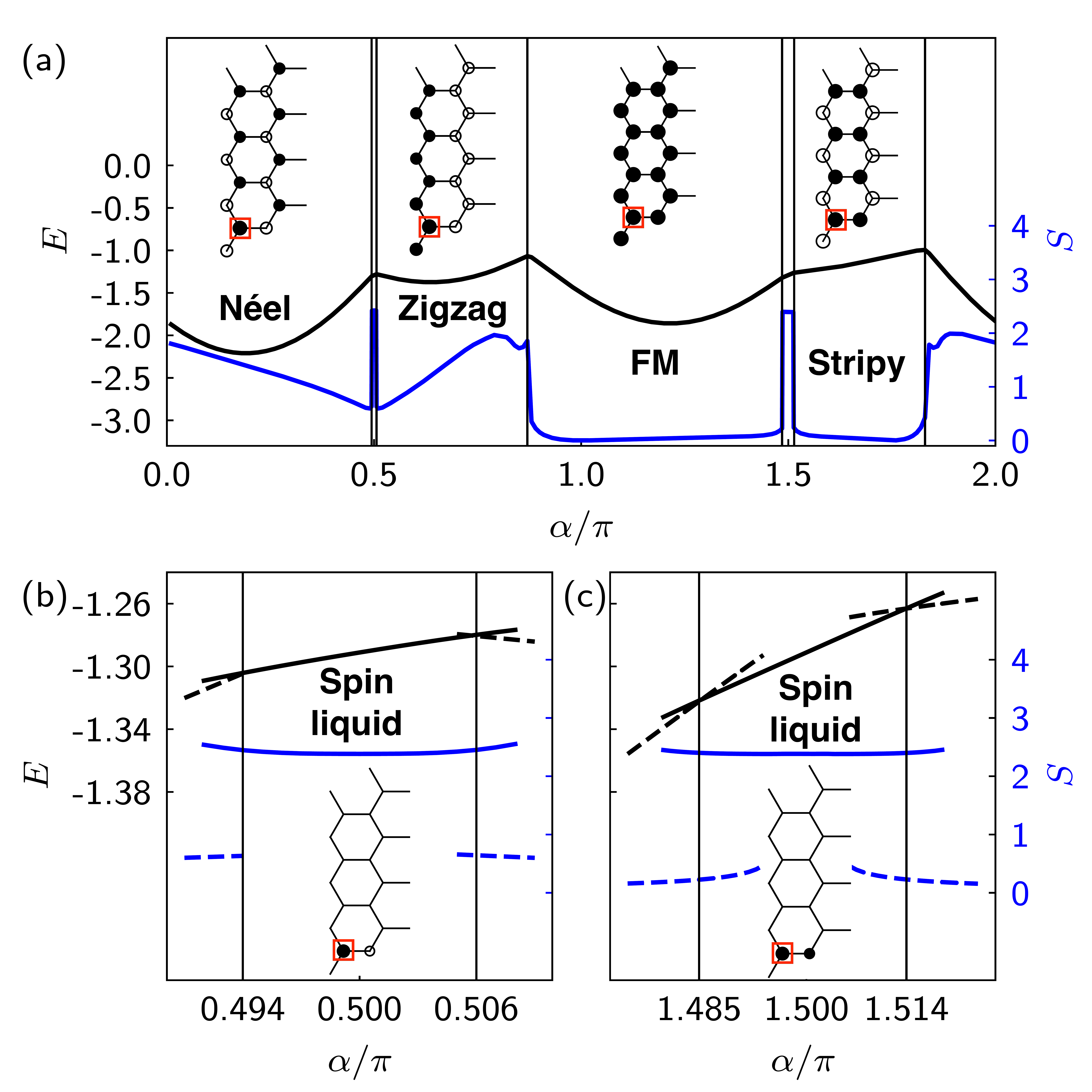}
 \caption{(a) Phase diagram of the spin-1 Kitaev-Heisenberg model with $\alpha\in[0,2\pi]$ on an infinite cylinder with $L_y=4$. The black line is the energy of the ground state per site, and the blue line is the entanglement entropy by cutting the cylinder along a ring. The bond dimension used in iDMRG is $\chi = 1000$. (b) and (c) are the zoom-in plots of the spin liquid phases around $\alpha=0.5\pi$ and $1.5\pi$, respectively. The black solid lines are the ground state energy per site with calculations initialized in the spin liquid phases, and the black dashed lines are the ground state energy per site with calculations initialized in the corresponding nearby symmetry broken phases. The blue lines are the entanglement entropy with similar meaning. The phase boundaries are determined by the discontinuous points of the first-order derivatives of the energy of the lowest energy state at each $\alpha$ with respect to $\alpha$. (insets) The spin-spin correlations $\langle S^x_iS^x_0\rangle$ of each phase in real space with $0$ sites denoted by the red squares. 
 \label{fig:phases}}
\end{figure}

\textbf{Phase diagram.}
We consider the spin-1 Kitaev-Heisenberg model on a 2D honeycomb lattice with the Hamiltonian:
\begin{eqnarray}
\hat{H} = K\sum_{\langle i,j\rangle_{\gamma}}\hat{S}^{\gamma}_i\hat{S}^{\gamma}_j + J\sum_{\langle i,j\rangle}\hat{\mathbf{S}}_i \cdot \hat{\mathbf{S}}_j 
\end{eqnarray} 
where $K = 2\sin(\alpha)$ and $J = \cos(\alpha)$ with $\alpha\in[0,2\pi]$ are the Kitaev and Heisenberg exchange couplings, respectively, $\hat{S}^{\gamma}_i$ is the $\gamma~ (=x, y, z)$ component of spin-1 operators on site $i$, and $\langle i,j\rangle_{\gamma}$ denotes the nearest neighbor coupled by $\gamma$-link.

We use an infinite DMRG (iDMRG) method~\cite{schollwock2011density,tenpy} on an infinite cylinder with circumference $L_y=4$, where $L_y$ is the number of unit cells of a honeycomb lattice around the cylinder, to get the phase diagram with respect to the parameter $\alpha$. The translational invariant building block of the infinite cylinder we use has two unit cells in $x$-directions along the cylinder to be compatible with the symmetry broken phases. Fig.~\ref{fig:phases}(a) shows the phase diagram with the ground state energy per site and the von Neumann entanglement entropy $S = - \sum_{\beta}\Lambda^2_{\beta}\log\Lambda^2_{\beta}$, where $\Lambda_{\beta}$s are the Schmidt values on the bond of matrix product state (MPS) cutting the infinite cylinder into two half along a ring. There are four magnetic ordered phases with spontaneous symmetry breaking: N\'{e}el phase for $\alpha$ in $[1.83\pi, 2\pi]$ and $[0, 0.494\pi]$, zigzag phase in $[0.506\pi,0.87\pi]$, ferromagnetic phase in $[0.87\pi, 1.485\pi]$ and stripy phase in $[1.514\pi, 1.83\pi]$, respectively. The insets of Fig.~\ref{fig:phases}(a) give the real-space correlation function $\langle \hat{S}^x_i\hat{S}^x_0\rangle$ for each ordered phase, with $0$ site marked by red squares. The black dot (circle) on each lattice site represents the positive (negative) sign of the correlation and the size of the dots and circles are proportional to their absolute values. The spin symmetry-breaking directions of the shown states are along the $x$-direction, i.e., $\langle \hat{S}^x_i\rangle\neq 0$. Therefore, the correlation $\langle \hat{S}^x_i\hat{S}^x_0\rangle$ has the longest correlation length and decays very slowly. At $\alpha=0.5\pi$ and $\alpha=1.5\pi$, the Heisenberg term $J=0$, thus the Hamiltonian is pure Kitaev with $K=2$ and $-2$, respectively. Two extended spin liquid phases exist around these pure Kitaev points in the region $[0.494\pi, 0.506\pi]$ and $[1.485\pi,1.514\pi]$ with details shown in Fig.~\ref{fig:phases}(b) and (c). In the spin liquids, the quantum frustration from Kitaev coupling terms dominate, thus no magnetic order appears. The static correlations are short-range to the nearest neighbors in the spin liquid as plotted in the insets of Fig.~\ref{fig:phases}(b) and (c). The signs of nearest neighbor correlation are opposite around $\alpha=0.5\pi$, and the same around $1.5\pi$, which is consisted with the corresponding anti-ferromagnetic and ferromagnetic Kitaev coupling. The regions of the spin liquid phases are smaller than that of the same model with spin-1/2 \cite{gohlke2017dynamics}.
The results of iDMRG indicate that the transitions between spin liquid and its nearby symmetry broken phases are first-order phase transitions. By initializing the calculation from each phase and moving towards the phase transition points, we could find a crossing in the energy lines, as shown by the solid and dashed black lines in Fig.~\ref{fig:phases}(b) and (c). It means that, for example, if we initialize the calculation with a state in the spin liquid, it will stay in the spin liquid phase even if the parameter of the model is in the symmetry broken phase near to the phase transition point, but the state we obtain in this way is a metastable state which has higher energy than the true ground state with symmetry broken. This is a key feature of the first-order phase transitions~\cite{motruk2017phase}. 
 
\begin{figure}[h]
 \includegraphics[width=\columnwidth]{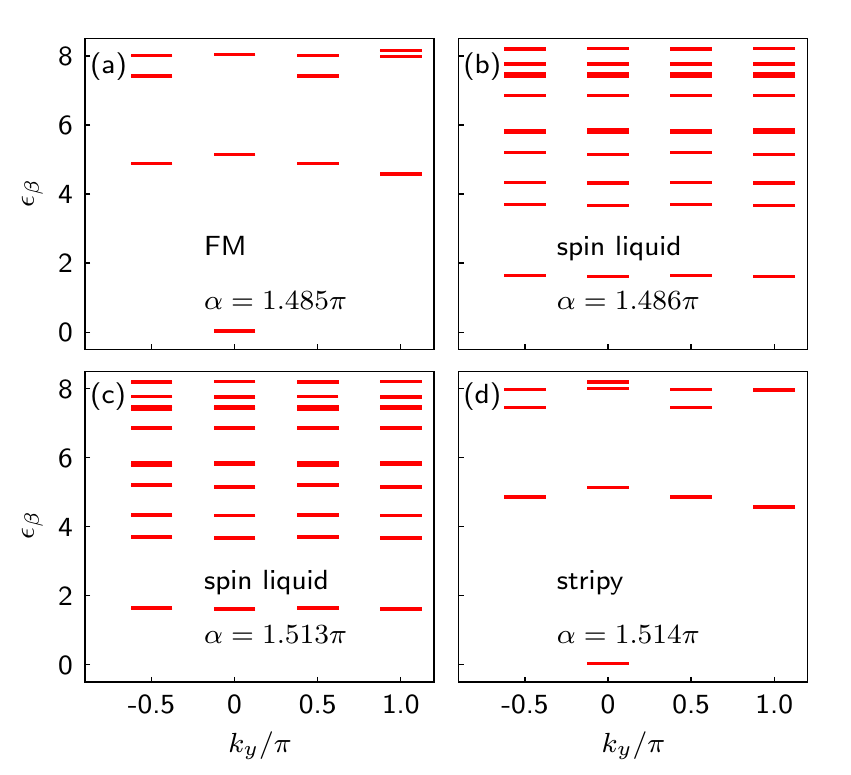}
 \caption{Entanglement spectrum $\epsilon_{\beta}$ with momentum $k_y$ around the cylinder of the ground states with (a) $\alpha=1.485\pi$ in ferromagnetic phase, (b) $\alpha=1.486\pi$ in spin liquid, (c) $\alpha=1.513\pi$ in spin liquid and (d) $\alpha=1.514\pi$ in stripy phase. All of these $\alpha$ points are the nearest points to the phase transitions in the order of $0.001\pi$.
 \label{fig:entanglement}}
\end{figure}

The entanglement spectrum also shows different structures in the spin liquid and magnetically ordered phases. The cylinder has translational symmetry along the $y$-direction, thus the Schmidt states are eigenstates of the momentum $k_y$ in the $y$-direction and the corresponding entanglement energy levels $\epsilon_{\beta} = -2\ln\Lambda_{\beta}$ can be labeled with $k_y$. The entanglement spectrum near the phase transition point from the ferromagnetic phase to spin liquid are plotted in Fig.~\ref{fig:entanglement}(a)(b), and from spin liquid to stripy magnetic ordered phase are given in Fig.~\ref{fig:entanglement}(c)(d). In the ordered phases, the entanglement spectrum has a non-degenerate dominate value, while in the spin liquid the entanglement spectrum has approximate four-fold degeneracies with different $k_y$. The abrupt changes of the spectrum structure at the phase transition points are consisted with the first-order phase transitions. At the pure Kitaev points ($K=2$), when the ground state is represented by an MPS with bond dimension $\chi=1000$, the four-fold degeneracy exists with the splitting in the order of 0.01 (the lowest four $\epsilon_{\beta}$ are  $[1.6672, 1.6676, 1.6676, 1.6681]$), while at the phase boundary as shown in Fig.~\ref{fig:entanglement}(b) the four-fold degeneracy splitting
increases to  the order of 0.1 (the lowest four $\epsilon_{\beta}$ are  $[1.6551, 1.6553, 1.6734, 1.6737]$). 
This degeneracy indicates that there is a non-trivial operator $\hat{O}$ that commutes with the entanglement Hamiltonian, which also gives rise to the degeneracy
in the edge spectrum.
We have checked that the spin liquid phases of spin-1/2 Kitaev model also have similar entanglement structures indicating strong superposition nature
of the spin liquid state.

\textbf{Pure Kitaev model and spin liquid.} In this part we will focus on the pure Kitaev model with $K = \pm 2$ and $J=0$, and the corresponding spin liquid phases. Let's now review some symmetry properties of the pure Kitaev model. The general Wilson loop operator for spin-$S$ on sites $\{1,2,...,n\}$ of a loop $L$ is defined by 
\begin{eqnarray}
\hat{W}_{L} &=& e^{i\pi \hat{S}^{\gamma_{12}}_{1}}e^{i\pi \hat{S}^{\gamma_{12}}_2}e^{i\pi \hat{S}^{\gamma_{23}}_{2}}e^{i\pi \hat{S}^{\gamma_{23}}_3}\cdots \nonumber\\
&&\cdot e^{i\pi \hat{S}^{\gamma_{(n-1)n}}_{(n-1)}}e^{i\pi \hat{S}^{\gamma_{(n-1)n}}_n}e^{i\pi \hat{S}^{\gamma_{n1}}_{n}}e^{i\pi \hat{S}^{\gamma_{n1}}_1}
\end{eqnarray}
where $\gamma_{i,i+1}(=x,y,z)$ denotes the directions of the link between site $i$ and $i+1$. It can be verified that $[\hat{W}_L,\hat{H}_K]=0$ and $[\hat{W}_L,\hat{W}_{L'}]=0$, where $\hat{H}_K$ denotes the pure Kitaev Hamiltonian. Since $(\hat{W}_L)^2=1$, the eigenvalues of $\hat{W}_L$ are $W_L = \pm 1$. It can be checked that this general definition reduces to the correct form in the spin-1/2 case. We rename $\hat{W}_L$ as $\hat{W}_p$ if the loop is around a single hexagonal plaquette of the honeycomb lattice ( $\hat{W}_p$ is also called as the $Z_2$ flux operator of a plaquette), and as $\hat{W}_y$ if the loop winds once around the cylinder.

To examine the nature of the spin liquid, we first check if the spin liquid is gapless. We choose two sizes of infinite cylinders with $L_y=3$ and $L_y=4$ and consider $\alpha=1.5\pi$ ($\alpha=0.5\pi$ is equivalent). We initialize the iDMRG calculations by complex random MPSs with bond dimension $\chi_{\rm{ini}}=100$. A random initial state may give a random bias to help  selecting states   in different sectors with different $W_y$ after updating variationally, thus we have a chance to get the lowest energy state in each sector. We name the state with a lower energy of these two states as the ground state and the other one as an excited state. 

\begin{figure}[ht]
 \includegraphics[width=\columnwidth]{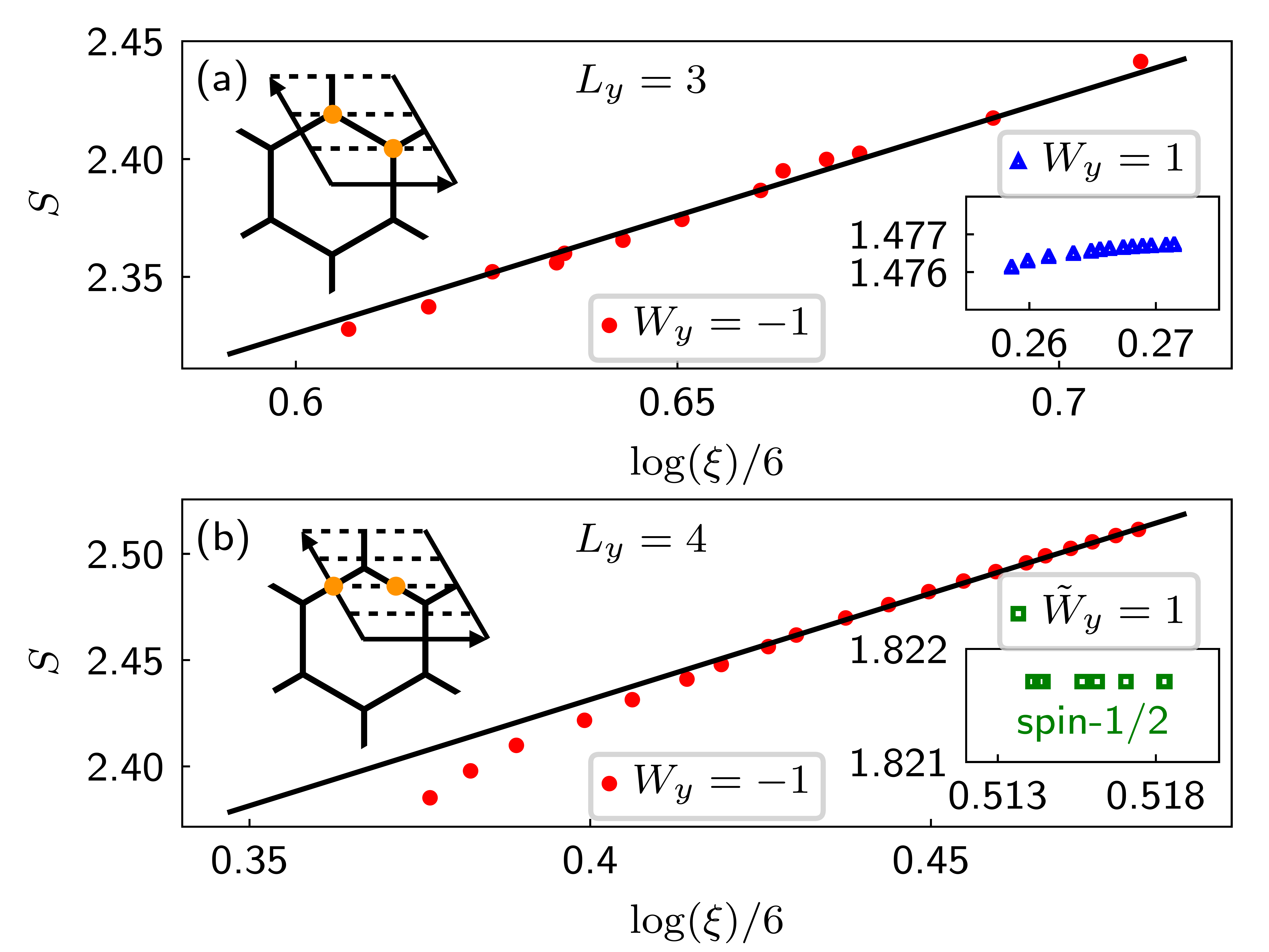}
 \caption{(a) Finite entanglement scaling of the lowest energy state with $W_y=-1$ on an infinite cylinder with $L_y=3$. The lower right inset is the same plot for the corresponding gapped ground state with $W_y=1$. (b) Finite entanglement scaling of the ground state with $W_y=-1$ on an infinite cylinder with $L_y=4$. The lower right inset is the same plot for the gapped ground state of spin-1/2 Kitaev model with $\tilde{W}_y=1$ on an infinite cylinder with $L_y=4$. The slope of the black lines in (a) and (b) give the central charge $c=1$. The upper left insets of (a) and (b) are the Brillouin zone of the honeycomb lattice with $L_y=3$ and $4$, respectively, in which the dashed lines are the momentum lines with periodic boundary condition in $y$-direction, and orange dots show the possible positions of gapless Majorana cones. 
 \label{fig:central_charge}} 
\end{figure}

For $L_y=3$, the ground state is gapped with $W_y=1$, while the lowest energy state with $W_y=-1$ (an excited state) is gapless. As an indication to a gapless state modeled from finite dimension MPSs, both the entanglement entropy $S$ and correlation length $\xi$ of the  variational-optimized MPS will increase with its bond dimension, and the scaling relation between them $S= (c/6)\log\xi$ gives the central charge $c$ of the gapless state. Fig.~\ref{fig:central_charge}(a) shows the finite entanglement scaling for this gapless excited state. The black linear line has the central charge $c=1$. The same plot for the gapped ground state is given in the lower right inset of Fig.~\ref{fig:central_charge}(a). The entanglement entropy of the gapped ground state converges quickly and does not grow with the bond dimension. This observation is quite similar to the spin-1/2 case. In spin-1/2 case, although with $L_y=3$ and periodic boundary condition in $y$-direction the momentum lines cut through the positions of the gapless Majorana cones at $\pm(\frac{1}{3},\frac{2}{3})$ 
in momentum space (see the upper left inset of Fig.~\ref{fig:central_charge}(a)), the free fermion degree of freedom will adjust to the anti-periodic boundary condition (due to the gauge flux) to give a gapped ground state with lower energy. 
While there is no  analytical solution in the spin-1 case, we  conjecture that similar
effect may also play an important role in spin-1 case as we have seen different flux state ($W_y=\pm 1$) have
different central charges  in finite size system. 
To identify the pattern of gapless models,  we consider $L_y=4$ in the following.

 For $L_y=4$, the ground state of spin-1 case is already gapless and have a central charge $c=1$ as shown in Fig.~\ref{fig:central_charge}(b). This gapless ground state has $W_y=-1$, which is the same as  the gapless excited state of $L_y=3$. The lowest energy state with $W_y=1$ is also gapless, but has a larger central charge. This is different from  the spin-1/2 case, in which $L_y=4$ is not compatible with the position of the Dirac cones, so we  always get a gapped ground state as shown in the lower right inset of Fig.~\ref{fig:central_charge}(b). 
One possible explanation of the central charge 1 here is that there are two gapless Majorana cones in the spin-1 case, and they may locate at the $(0,\frac{1}{2})$ and $(\frac{1}{2},\frac{1}{2})$ as marked by the orange dots in the upper left inset of Fig.~\ref{fig:central_charge}(b). We could see that with $L_y=4$ and periodic boundary condition in $y$-direction the momentum lines go through these two points, and give the gapless ground state with central charge $c=1$ (one Majorana cone contributes $1/2$ to the central charge). Notice that the $C_6$ rotation symmetry is broken by the cylinder geometry, thus the spectrum at $(\frac{1}{2},0)$ may be different from that at $(0,\frac{1}{2})$ or $(\frac{1}{2},\frac{1}{2})$. Other  possible situations
include gapless excitations 
around  $\Gamma=(0,0)$ point, which  is protected by some symmetries,   or  around a small Fermi-Sea, which are consistent
with  both $L_y=3$ and $L_y=4$ results.


\begin{figure}[ht]
 \includegraphics[width=\columnwidth]{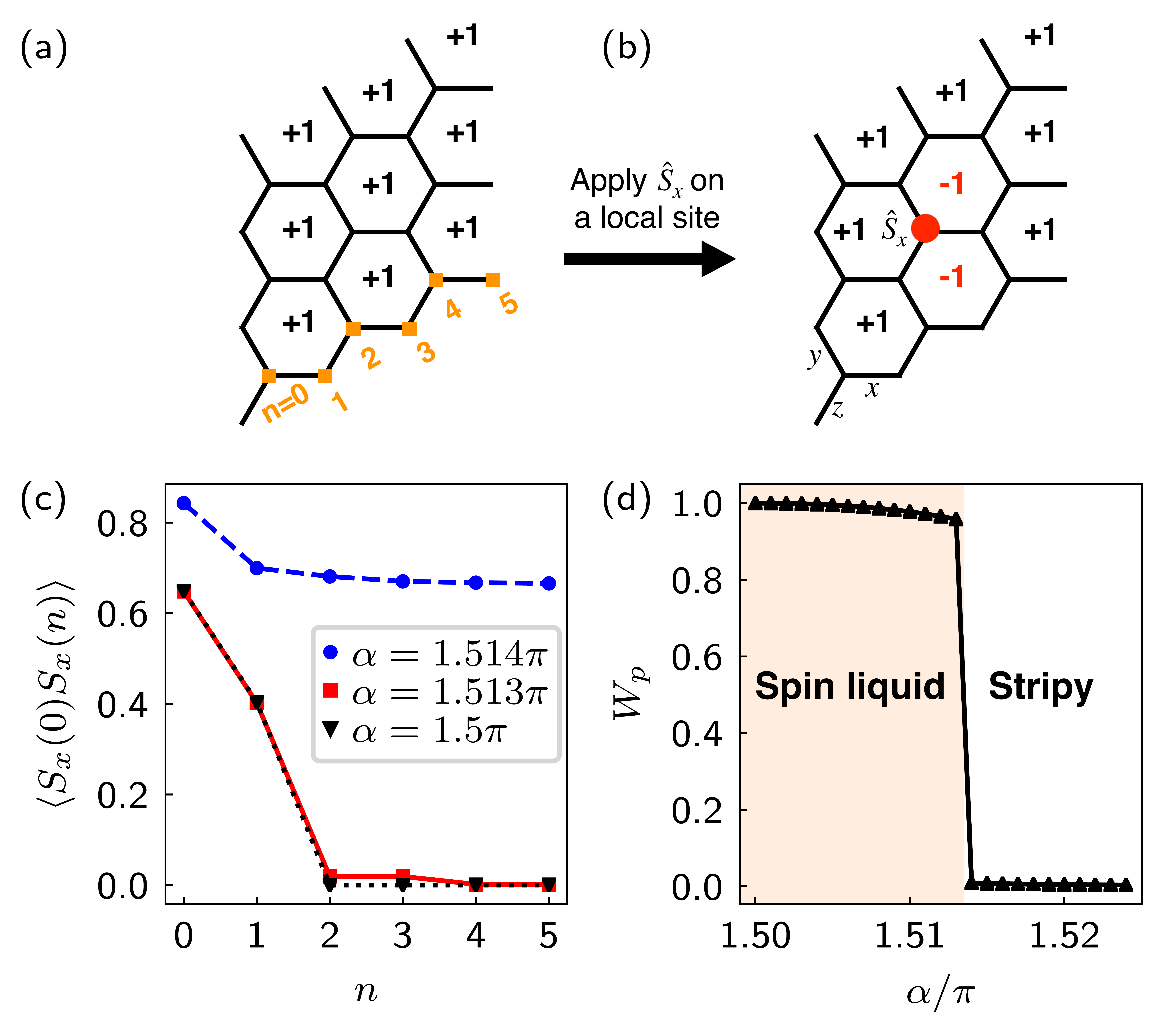} 
 \caption{(a) The real space lattice with flux $+1$ in each plaquette is schematically shown for the ground state of pure Kitaev model. The sites $n=0,1,...,5$ with orange squares are used to calculate correlations in (c). (b) Apply $\hat{S}_x$ on one site of the ground state in (a), flux in two plaquettes are flipped to $-1$. (c) The spin-spin correlations $\langle S^x_0S^x_n\rangle$ on $n$ labelled in (a) are plotted at $\alpha = 1.5\pi, 1.513\pi,$ and $1.514\pi$. (d) The flux $W_p$ is each plaquette versus $\alpha$ around the phase transition between spin liquid and stripy phase.
 \label{fig:flux}}
\end{figure}

Beside $\hat{W}_y$, the eigenvalue of local flux operator $\hat{W}_p$ is also a good quantum number of the eigenstates of the pure Kitaev Hamiltonian. The ground states we obtained is vortex-free, i.e., eigenvalues of $\hat{W}_p$ are $+1$ for all of the plaquettes. The flux $W_p$ in each plaquette is near to 1 in the spin liquid phase with $J\neq 0$ around the pure Kitaev points, and drops abruptly to near 0 at the phase transition point as shown in Fig.~\ref{fig:flux}(d).

By direct calculation, we know that $\{\hat{S}^{\gamma}_i, \hat{W}_p\}=0$ if site $i$ is on the plaquette $p$ and $\gamma$ corresponds to one of the two links on $p$ which is connected to site $i$. On the other hand,   $[\hat{S}^{\gamma'}_j, \hat{W}_p]=0$ if site $j$ has no overlap with the plaquette $p$ or $\gamma'$ does not correspond to the two links  on $p$
 connecting  to site $j$.     
Therefore, if we apply a local spin operator, for example, $\hat{S}^x_i$ on site $i$ of the lattice, two of the plaquettes which share the site $i$ and link $x$ will reverse their flux $W_p$ as schematically shown in Fig.~\ref{fig:flux}(a) and (b). This property leads to the vanishing of the spin-spin correlations beyond the nearest neighbors, since the eigenstates with different quantum number $W_p$ will be orthogonal to each other and thus have zero overlap. This correlation remains short-ranged in the entire spin liquid phase, and changes to long-range once the symmetry-breaking happens. In Fig.~\ref{fig:flux}(c), the spin-spin correlation $\langle S^x_0S^x_n\rangle$ is plotted with sites $n=0,1,\dots, 5$ as marked in Fig.~\ref{fig:flux}(a), in which $\alpha=1.5\pi$ is a pure Kitaev point, $\alpha=1.513\pi$ and $\alpha=1.514\pi$ are adjacent to the phase transition point between spin liquid and stripy phase. We could see that the change of the spin-spin correlation is very sharp crossing the phase transition.

\textbf{Discussion.} We identify the gapless spin liquid phases in spin-1 Kitaev-Heisenberg model, and find that the gapless modes
are  very different from the corresponding spin-1/2 model. The extensive numerical studies in this work give some insight  about the nature
of the gapless excitations and could motivate more theoretical studies to fully understand the spin-1 Kitaev model.
There are still many open questions about extended spin-1 Kitaev model for future works. We list a few here: (1) If the Kitaev coupling coefficients are anisotropic, is there a gapped region and what is the  topological nature  of the gapped phase? Since the local Hilbert space has a dimension three, it may go beyond the physics of the $Z_2$ toric code. (2) If we apply a magnetic field, will the spin liquid open a gap and become a non-Abelian phase? Is it still the Ising type non-Abelian phase? (3) The real materials which realize spin-1 Kitaev coupling may have many other complicated interactions like the spin-1/2 case, so the phase diagram could be much richer. Therefore, more theoretical studies with different coupling  terms added to the pure Kitaev model are
also interesting. 

Note added: While completing this manuscript, we became aware of a related work \cite{Lee2019tensor} based
on  tensor network approach  for spin-1 Kitaev model.  Some conclusions 
 of their work on the nature of spin liquid does not agree with our work, and we believe that further theoretical and numerical studies are demanded to resolve the full nature of the spin liquid.

\textbf{Acknowledgements} 
We thank Hong-Hao Tu, Matthias Gohlke, Rong-Yang Sun, Wayne Zheng and Zheng Zhu  for helpful discussions. iDMRG calculations in this paper were performed using the TeNPy Library (version 0.3.0)~\cite{tenpy}. Work by Xiao-Yu Dong and D. N. Sheng were supported by the Department of Energy, Office of Basic Energy Sciences, Division of Materials Sciences and Engineering, under Contract No. DE-AC02- 76SF00515 through SLAC National Accelerator Laboratory.
\bibliography{Kitaev}
\end{document}